\documentclass[prb,aps,twocolumn,superscriptaddress]{revtex4}
\usepackage{graphicx,amsmath,bm}
\usepackage{dcolumn}

\begin{document}

\title{Studying Superfluid Transition of a Dilute Bose Gas \\
by Conserving Approximations}

\author{Kazumasa T{\sc sutsui} and Takafumi K{\sc ita}}
%\email{}
\affiliation{Department of Physics, Hokkaido University, Sapporo 060-0810, Japan}
\date{\today}

\begin{abstract}
We consider the Bose-Einstein transition of homogeneous weakly interacting spin-0 particles based 
on the normal-state $\Phi$-derivable approximation.
Self-consistent calculations of Green's function and the chemical potential with several approximate $\Phi$'s are performed numerically as a function of temperature near $T_{\rm c}$, which exhibit qualitatively different results.
The ladder approximation apparently shows a continuous transition with the prefactor $c=2.94$ for the transition-temperature shift 
$\Delta T_{\rm c}/T_{\rm c}^{0}=ca n^{1/3}$ given in terms of the scattering length $a$ and density $n$.
In contrast, the second-order, particle-hole, and fluctuation-exchange approximations yield a first-order transition. 
The fact that some standard $\Phi$'s predict a first-order transition challenges us to clarify whether or not
the transition is really continuous.
\end{abstract}

\maketitle

\section{Introduction}

The Bose-Einstein condensation (BEC) 
of homogeneous weakly interacting Bose gases has attracted much attention over a decade.\cite{GCL97,HGL99,Baym99,Baym99-2,Holzmann99,Baym00,Arnold00,AM01,KPS01,CPRS02,Kleinert03,Kastening03,Andersen04,LHK04,NL04,BGW06,PGP08}
As shown by Baym {\em et al}.,\cite{Baym99,Baym99-2} this topic is profound enough to require treatments beyond the simple perturbation expansion.
To be specific, they confirmed that the transition temperature $T_{\rm c}$ starts to increase linearly with the $s$-wave scattering length $a$ as
\begin{equation}
\Delta T_{\rm c} /T_{\rm c}^{0}=can^{1/3},
\label{DeltaTc}
\end{equation}
where $n$ is the density, and presented analytic estimates for the prefactor $c$ using various approximations for Green's function.
Subsequently, a couple of Monte Carlo simulations on finite lattices obtained a widely accepted value $c\approx 1.3$.\cite{AM01,KPS01}

However, these studies as well as others\cite{GCL97,HGL99,Baym99,Baym99-2,Holzmann99,Baym00,Arnold00,AM01,KPS01,CPRS02,Kleinert03,Kastening03,Andersen04,LHK04,NL04,BGW06,PGP08}
 focused mostly on the critical density by implicitly assuming a continuous transition,
thereby leaving behind an important question of how the system approaches the critical point as a function of temperature.

We will consider the issue based on the conserving $\Phi$-derivable approximation.\cite{LW60,Baym62,BS89,Kita10b} 
This systematic approximation scheme has several remarkable advantages,\cite{Kita10b} 
as may be realized by the fact that the Bardeen-Cooper-Scheriffer theory of superconductivity belongs to 
it as a lowest-order approximation,\cite{Kita11b}
and has been used extensively to clarify anomalous properties of high-$T_{\rm c}$ cuprate superconductors.\cite{BS89}
Thus, the method will help us to see the critical region of BEC more closely. 
Indeed, the contents here may be regarded as an extension of those with self-consistent approximations 
by Baym {\em et al}.\  \cite{Baym99,Baym99-2} just on $T=T_{\rm c}$ to 
(i) incorporate temperature dependences of $T\gtrsim T_{\rm c}$ and 
(ii) consider more approximations systematically.
We will thereby find that the self-consistent one bubble approximation they considered 
yields a first-order transition contrary to their assumption.
The main results are summarized in \S3.2 below.

\section{Formulation}

\subsection{Hamiltonian and Green's function}

We will consider identical homogeneous bosons with spin $0$, mass $m$, and density $n$ interacting via a weak contact potential
$U\delta({\bf r}_{1}-{\bf r}_{2})$. To study this system near $T_{\rm c}$, 
we adopt the units
\begin{equation}
m=\frac{1}{2},\hspace{5mm}n=\frac{\zeta(3/2)}{(4\pi)^{3/2}}, \hspace{5mm} k_{\rm B}=\hbar=1,
\label{units}
\end{equation}
where $\zeta(3/2)\!=\! 2.612\cdots$ is the Riemann zeta function and $k_{\rm B}$ denotes the Boltzmann constant.
Thus, the critical temperature of the ideal Bose gas,\cite{AL03,PS08} 
$T_{\rm c}^{0}=(2\pi\hbar^2/k_{\rm B}m)[n/\zeta(3/2)]^{2/3}$, is set equal to $1$, 
and the kinetic energy is expressed in terms of the momentum ${\bm p}$ simply as $\epsilon_{\bm p}=p^2$.

The Hamiltonian is given by
\begin{equation}
H = \sum_{\bm p} (\epsilon_{\bm p}-\mu)c_{\bm p}^{\dagger} c_{\bm p} + \frac{U}{2}\sum_{{\bm p}_1{\bm p}_2{\bm q}}c_{{\bm p}_1+{\bm q}}^{\dagger} 
c_{{\bm p}_2-{\bm q}}^{\dagger} c_{{\bm p}_2}c_{{\bm p}_1} ,
\end{equation}
where $\mu$ is the chemical potential
and $c_{\bm p}^{\dagger}$ and $c_{\bm p}$ are the creation and annihilation operators, respectively. 
Ultraviolet divergences inherent in the continuum model are cured here
by introducing a momentum cutoff $p_{\rm c}\gg 1$. 
However, our final results will be free from $p_{\rm c}$, as seen below.
It is standard in the low-density limit to remove $U$ in favor of the $s$-wave scattering length $a$.
They are connected in the conventional units by\cite{PS08}
$$
\frac{m}{4\pi \hbar^2 a}= \frac{1}{U}+\int\frac{{\rm d}^3 p}{(2\pi\hbar)^3} \frac{\theta (p_{\rm c}-p)}{2\epsilon_{\bm p}} 
$$
with $\theta (x)$ the step function, which in the present units reads
$1/8\pi a =1/U+p_{\rm c}/4\pi^2$. 
We will focus on the limit $a \rightarrow  0$ and choose $p_{\rm c}$ so that $1\ll p_{\rm c}\ll \pi/2a$ is satisfied.
Thus, we can set
\begin{equation}
U=8\pi a
\label{U-a}
\end{equation}
to an excellent approximation.

Let us introduce Green's function in the normal state by
\begin{subequations}
\begin{equation}
G(\vec{p}\,)=\frac{1}{i\varepsilon_n-\epsilon_{\bm p}-\Sigma(\vec{p}\,)+\mu} ,
\label{Dyson}
\end{equation}
where $\vec{p}\equiv ({\bm p},i\varepsilon_{n})$ with $\varepsilon_n\!=\! 2n\pi T$ ($n\!=\! 0,\pm 1,\pm 2,\cdots$) 
a boson Matsubara frequency.
The self-energy $\Sigma(\vec{p}\,)$ is given exactly by
\begin{equation}
\Sigma(\vec{p}\,) =-\frac{1}{T} \frac{\delta \Phi}{\delta G(\vec{p}\,)} ,
\label{Sigma-def}
\end{equation}
\end{subequations}
where the functional $\Phi=\Phi[G]$ is defined as the infinite sum of closed skeleton diagrams
in the simple perturbation expansion for the thermodynamic potential with the replacement $G_0\rightarrow G$.\cite{LW60,Baym62,Kita10b}
The chemical potential $\mu$ is connected with the particle density $n$ by
\begin{equation}
n=-\frac{T}{{\cal V}}\sum_{\vec{p}}G(\vec{p}\,)\, {\rm e}^{i\varepsilon_n 0_+},
\label{n-mu}
\end{equation}
with ${\cal V}$ the volume and $0_+$ an infinitesimal positive constant.

If BEC is realized as a continuous transition, the transition temperature $T_{\rm c}$ will be determined by 
the condition:
\begin{equation}
\mu = \Sigma(\vec{0}).
\label{Tc-condition}
\end{equation}
This relation, which is derived from the Hugenholtz-Pines relation in the condensed phase \cite{Kita09,HP59}
by setting the off-diagonal self-energy zero, naturally extends the condition $\mu_0(1)\!=\! 0$ for the  
transition temperature of the ideal gas\cite{PS08}
to interacting cases. Indeed, it was used by Baym {\em et al}.\ \cite{Baym99,Baym99-2} to estimate the critical density $n_{\rm c}$.

\begin{figure}[tb]
\begin{center}
\includegraphics[width=0.8\linewidth]{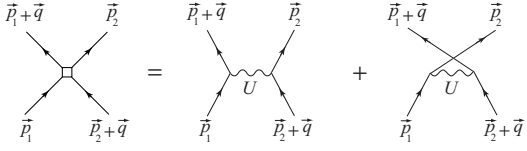}
\caption{Diagrammatic representation of symmetrized vertex $\Gamma^{(0)}(\vec{p}_1,\vec{p}_2,\vec{q})$,
which is equal to $2U$ for the contact interaction $U\delta({\bm r}_1-{\bm r}_2)$ used here.}
\label{Fig1}
\end{center}
\end{figure}

\subsection{FLEX and related approximations}

The $\Phi$-derivable approximation consists of (i) retaining some finite terms or partial series in $\Phi[G]$ and 
calculating $G$ and $\Sigma$ self-consistently by eqs.\ (\ref{Dyson}) and (\ref{Sigma-def}).
We will consider the fluctuation-exchange (FLEX) approximation\cite{BS89} and those derivable from it by reducing terms. 
All of them are concisely expressible
in terms of the symmetrized vertex $\Gamma^{(0)}(\vec{p}_1,\vec{p}_2,\vec{q})$ of Fig.\ \ref{Fig1},
which is equal to $2U$ with no momentum and frequency dependence for the present contact interaction 
$U\delta({\bm r}_1-{\bm r}_2)$.\cite{AGD63}

\begin{figure}[tb]
\begin{center}
\includegraphics[width=0.9\linewidth]{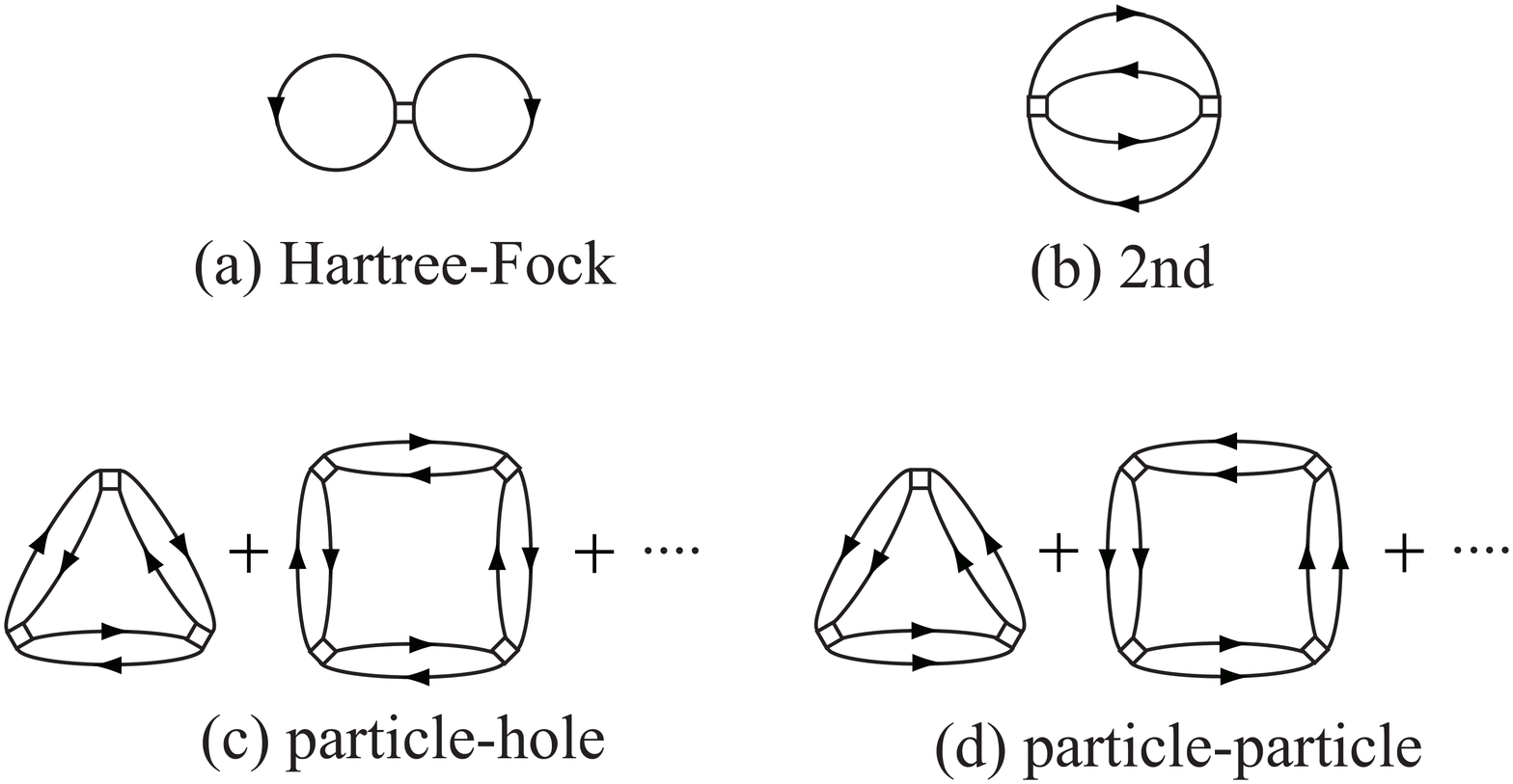}
\caption{Feynman diagrams for $\Phi_{\rm FLEX}$
\label{Fig2}}
\end{center}
\end{figure}

To begin with, $\Phi_{\rm FLEX}$ is given as a sum of four kinds of diagrams in Fig.\ \ref{Fig2} as
\begin{equation}
\Phi_{\rm FLEX}=\Phi_{1}+\Phi_2+\Phi_{\rm ph}+\Phi_{\rm pp},
\label{Phi_FLEX}
\end{equation}
where $\Phi_{1}\equiv {\cal V}Un^2$ is the  Hartree-Fock term.
To express the other contributions analytically, let us introduce the functions
\begin{subequations}
\label{chi_ph-pp}
\begin{equation}
\chi_{\rm ph}(\vec{q}\,)\equiv\frac{T}{{\cal V}}\sum_{\vec{p}}G(\vec{p}\,)G(\vec{p}+\vec{q}\,),
\end{equation}
\begin{equation}
\chi_{\rm pp}(\vec{q}\,)\equiv\frac{T}{{\cal V}}\sum_{\vec{p}}G(\vec{p}\,)G(-\vec{p}+\vec{q}\,),
\end{equation}
\end{subequations}
each of which corresponds to a pair of lines connecting adjacent vertices in Fig.\ \ref{Fig2}(c) and (d), respectively.
By following the Feynman rules for the perturbation expansion in terms of $\Gamma^{(0)}\!=\! 2U$,\cite{Kita11b}
the numerical factor of  Fig.\ \ref{Fig2}(b) and those of the $n$th-order  diagrams
($n=3,4,\cdots$) in Fig.\ \ref{Fig2}(c) and (d) are easily found as $c_2=-(-2U)^2 2^2T /4^2 2!=-U^2 T/2$,
$c^{(n)}_{\rm ph}=-(-2U)^n 2^{2n-1}(n-1)! T /4^n n!=(-2)^{n-1}U^n T/n$, and 
$c^{(n)}_{\rm pp}=-(-2U)^n 2^{n}(n-1)! T /4^n n!=(-1)^{n-1}U^n T/n$.
We thereby obtain analytic expressions for Fig.\ \ref{Fig2}(b)-(d) as
\begin{subequations}
\begin{equation}
\Phi_2=-\frac{T}{2}\sum_{\vec{q}}[U\chi_{\rm ph}(\vec{q}\,)]^2=
-\frac{T}{2}\sum_{\vec{q}}[U\chi_{\rm pp}(\vec{q}\,)]^2,
\label{Phi_2}
\end{equation}
\begin{eqnarray}
&&\hspace{-10mm}
\Phi_{\rm ph}=
\frac{T}{2}\sum_{\vec{q}}\biggl\{ \ln[1+2U\chi_{\rm ph}(\vec{q}\,)]
-2U\chi_{\rm ph}(\vec{q}\,)
\nonumber \\
&&\hspace{0.5mm}
+\frac{[2U\chi_{\rm ph}(\vec{q}\,)]^2}{2}\biggr\},
\label{Phi_ph}
\end{eqnarray}
\begin{eqnarray}
&&\hspace{-10mm}
\Phi_{\rm pp}=
T\sum_{\vec{q}}\biggl\{ \ln[1+U\chi_{\rm pp}(\vec{q}\,)]
-U\chi_{\rm pp}(\vec{q}\,)
\nonumber \\
&&\hspace{0.5mm}
+\frac{[U\chi_{\rm pp}(\vec{q}\,)]^2}{2}\biggr\} ,
\label{Phi_pp}
\end{eqnarray}
\end{subequations}
respectively.

Let us substitute eq.\ (\ref{Phi_FLEX}) into $\Phi$ of eq.\ (\ref{Sigma-def}). We then obtain the self-energy as
\begin{eqnarray}
&&\hspace{-10mm}
\Sigma(\vec{p})=2Un -\frac{T}{{\cal V}}\sum_{\vec{q}}[V_2(\vec{q})+V_{\rm ph}(\vec{q})]
G(\vec{p}+\vec{q})
\nonumber \\
&&\hspace{2mm}
-\frac{T}{{\cal V}}\sum_{\vec{q}}V_{\rm pp}(\vec{q})G(-\vec{p}+\vec{q}),
\label{Sigma_FLEX}
\end{eqnarray}
with
\begin{subequations}
\begin{equation}
V_{2}(\vec{q})\equiv -2U^2 \chi_{\rm ph}(\vec{q}),
\end{equation}
\begin{equation}
V_{\rm ph}(\vec{q})\equiv 2U \frac{[2U\chi_{\rm ph}(\vec{q})]^2}{1+2U\chi_{\rm ph}(\vec{q})},
\end{equation}
\begin{equation}
V_{\rm pp}(\vec{q})\equiv 2U \frac{[U\chi_{\rm pp}(\vec{q})]^2}{1+U\chi_{\rm pp}(\vec{q})}.
\end{equation}
\end{subequations}
Equation (\ref{Dyson}) with eq.\ (\ref{Sigma_FLEX}) forms a closed nonlinear equation
for $G(\vec{p})$ that may be solved numerically 
to clarify normal-state properties of $T\geq T_{\rm c}$.

Besides eq.\ (\ref{Sigma_FLEX}), we will also consider the following self-energies:
\begin{subequations}
\label{Sigma's}
\begin{equation}
\Sigma_{2}(\vec{p})\equiv 2Un -\frac{T}{{\cal V}}\sum_{\vec{q}}V_2(\vec{q})
G(\vec{p}+\vec{q}),
\label{Sigma_2}
\end{equation}
\begin{equation}
\Sigma_{\rm ph}(\vec{p})\equiv 2Un -\frac{T}{{\cal V}}\sum_{\vec{q}}[V_2(\vec{q})+V_{\rm ph}(\vec{q})]G(\vec{p}+\vec{q}),
\label{Sigma_ph}
\end{equation}
\begin{eqnarray}
&&\hspace{-10mm}
\Sigma_{\rm pp}(\vec{p})\equiv 2Un -\frac{T}{{\cal V}}\sum_{\vec{q}}[V_2(\vec{q})G(\vec{p}+\vec{q})
\nonumber \\
&&\hspace{5mm} +V_{\rm pp}(\vec{q})G(-\vec{p}+\vec{q})],
\label{Sigma_pp}
\end{eqnarray}
\end{subequations}
which are all derivable from eq.\ (\ref{Sigma_FLEX}) by reducing terms.
Using these different approximations, we may check how well the present approach describes the BEC transition.

Equations (\ref{Sigma_2}) and (\ref{Sigma_pp}) were used by Baym {\em et al}.\ \cite{Baym99,Baym99-2} 
as ``self-consistent one bubble approximation'' and 
``self-consistent ladder sum'' for estimating $T_{\rm c}$.
They also considered the ``self-consistent bubble sum'' of setting $V_{\rm ph}(\vec{q})\rightarrow \frac{1}{2}V_{\rm ph}(\vec{q})$
in eq.\ (\ref{Sigma_ph}), which amounts to neglecting the exchange processes altogether in the particle-hole series of Fig.\ \ref{Fig2}(c).
Since the exchange processes are definitely present, we will consider eq.\ (\ref{Sigma_ph}) instead of the self-consistent bubble sum.

\subsection{Dilute gas near $T_{\rm c}$}

We now focus on the critical region of the weak-coupling limit $a\rightarrow 0$, i.e.,  $T\gtrsim 1$.
Noting that this system is quantitatively close to the ideal gas, we first express the chemical potential as
\begin{equation}
\mu =\mu_0+\Sigma(\vec{0})+\Delta\mu ,
\label{mu=mu_0+dmu}
\end{equation}
where $\mu_0$ is the chemical potential of the ideal gas vanishing quadratically for $T\rightarrow 1$ as\cite{AL03}
\begin{equation}
\mu_0(T)  = -c_0^2 (T-1)^2, \hspace{5mm}c_0\equiv \frac{3\zeta(3/2)}{4\sqrt{\pi}}.
\label{mu_0}
\end{equation}
We also adopt the classical-field approximation by Baym {\em et al}.\ \cite{Baym99,Baym99-2} 
of retaining only the $\varepsilon_n=0$ component in eq.\ (\ref{Dyson}); its validity will be confirmed shortly.
We subsequently perform a change of variables, $T\rightarrow \tau$, $p\rightarrow \tilde{p}$, $\Delta\mu\rightarrow \Gamma$,
and $\Sigma({\bm p},i\varepsilon_n=0) \rightarrow  \sigma(\tilde{p})$, given explicitly by
\begin{subequations}
\label{COV}
\begin{equation}
T = 1+\frac{U}{c_0} \tau,
\label{T-tau}
\end{equation}
\begin{equation}
p = U \tau \tilde{p},
\label{p-tp}
\end{equation}
\begin{equation}
\mu_0(T)+\Delta\mu(T) = -(U\tau)^2 \Gamma(\tau),
\label{Gamma-def}
\end{equation}
\begin{equation}
\Sigma({\bm p},i\varepsilon_n=0) = \Sigma(\vec{0})+(U\tau)^2 \sigma(\tilde{p}).
\label{Sigma-sigma}
\end{equation}
\end{subequations}
Note that $\Gamma\rightarrow 1$ and $\sigma(p)\rightarrow 0$ for $U\rightarrow 0$ as seen from
eqs.\ (\ref{mu_0}) and (\ref{COV}).
The change of variables also removes the remaining source of the ultraviolet divergence, i.e., 
$\Sigma(\vec{0})$, completely from the theory.
Substituting eq.\ (\ref{mu=mu_0+dmu}) into eq.\ (\ref{Dyson}) for $\varepsilon_n=0$ 
and using eq.\ (\ref{COV}),
we can write $G({\bm p},i\varepsilon_n=0)$ as
\begin{equation}
G({\bm p},0)=-\frac{g(\tilde{p})}{(U\tau)^2} , \hspace{5mm}g(p)\equiv \frac{1}{p^2+\sigma(p)+\Gamma} .
\label{G-CFA}
\end{equation}
We also adopt the classical-field approximation for eq.\ (\ref{chi_ph-pp}) 
and substitute eqs.\ (\ref{T-tau}), (\ref{p-tp}), and (\ref{G-CFA}) into it.
It then turns out that
$\chi_{\rm ph}({\bm q},0)=\chi_{\rm pp}({\bm q},0)=\chi(\tilde{q})/U\tau$ with
\begin{equation}
\chi(q)\equiv \int\frac{{\rm d}^3 p}{(2\pi)^3}g(p)g(|{\bm p}+{\bm q}|) .
\label{chi}
\end{equation}
Subsequently using eqs.\ (\ref{Sigma_FLEX}) and (\ref{Sigma-sigma}) 
with eqs.\ (\ref{T-tau}), (\ref{p-tp}), and (\ref{G-CFA}),
we obtain an equation for the reduced self-energy $\sigma(p)$ as
\begin{equation}
\sigma(p)=\int\frac{{\rm d}^{3}q}{(2\pi)^3} v(q)[g(|{\bm p}+{\bm q}|)-g(q)],
\label{sigma}
\end{equation}
with
\begin{equation}
v(q)\equiv
\frac{2}{\tau} \left\{ -\frac{\chi(q)}{\tau}+ \frac{[2\chi(q)/\tau]^2}{1+2\chi(q)/\tau}+\frac{[\chi(q)/\tau]^2}{1+\chi(q)/\tau}\right\} .
\label{v(q)}
\end{equation}
As for eq.\ (\ref{n-mu}) for the chemical potential, we subtract its non-interacting correspondent from it, 
adopt the classical-field approximation, and use eq.\ (\ref{G-CFA}).
It is thereby transformed into an equation for $\Gamma=\Gamma(\tau)$ as
\begin{equation}
\int_0^\infty  \!\left[\frac{1}{p^2+\sigma(p)+\Gamma}-\frac{1}{p^2+1}\right] p^2 {\rm d} p =0 .
\label{Gamma-eq}
\end{equation}
Equations (\ref{sigma}) and (\ref{Gamma-eq}) form closed equations for $\sigma(p)$ and $\Gamma$ for a given $\tau$.
Note that the transformation (\ref{COV}) has removed $U$ completely from the self-consistent equations.
Besides, they are free from ultraviolet divergences.

The self-energies of eq.\ (\ref{Sigma's}) can be transformed similarly,
which turn out to have the kernels
\begin{subequations}
\label{v(q)-add}
\begin{equation}
v_2(q)\equiv
 -\frac{2\chi(q)}{\tau^2} ,
\label{v_2(q)}
\end{equation}
\begin{equation}
v_{\rm ph}(q)\equiv -\frac{2\chi(q)}{\tau^2}\,
\frac{1-2\chi(q)/\tau}{1+2\chi(q)/\tau} ,
\label{v_ph(q)}
\end{equation}
\begin{equation}
v_{\rm pp}(q)\equiv -\frac{2\chi(q)}{\tau^2} \,\frac{1}{1+\chi(q)/\tau} ,
\label{v_pp(q)}
\end{equation}
\end{subequations}
in place of $v(q)$ in eq.\ (\ref{sigma}). 
Note that $v_2(q)$ and $v_{\rm pp}(q)$ are negative, as seen from eq.\ (\ref{chi}), whereas 
$v(q)$ and $v_{\rm ph}(q)$ change sign from negative to positive as $\tau$ is decreased towards zero.
These distinct behaviors of the kernels from different approximations will lead to contradictory predictions on the BEC transition, as seen below.

Finally, eq.\ (\ref{Tc-condition}) for the continuous transition point is transformed with eqs.\ (\ref{mu=mu_0+dmu})
and (\ref{Gamma-def}) into
\begin{equation}
\Gamma(\tau_{\rm c})=0.
\label{Gamma(tau_c)=0}
\end{equation}
The corresponding critical temperature is easily obtained by eq.\ (\ref{T-tau}) with $\tau= \tau_{\rm c}$.
Using eqs.\ (\ref{units}), (\ref{U-a}), and (\ref{mu_0}) as well as $T_{\rm c}^{0}\!=\! 1$ in the present units,
we confirm that the transition-temperature shift starts linearly in $a$ as eq.\ (\ref{DeltaTc}) with the prefactor
\begin{equation}
c = \frac{(8\pi)^2}{3[\zeta(3/2)]^{4/3}}\tau_{\rm c} =58.52 \tau_{\rm c}.
\label{c-expression}
\end{equation}

A couple of comments are in order before closing the subsection.
First, eq.\ (\ref{sigma}) tells us that diagrams from second through infinite orders in $U$ contribute equivalently 
to $\sigma(p)$.
The validity of this statement is clearly not restricted to the FLEX approximation alone; it can be confirmed easily by
applying eqs.\ (\ref{COV}) and (\ref{G-CFA}) for general $n$th order terms in the classical-field approximation.
Thus, we need to include infinite diagrams of $\Phi$ to obtain an exact value of $\tau_{\rm c}$ in the self-consistent perturbation approach, which is practically impossible. We may expect, however, that some approximations for $\Phi$  enable us to obtain qualitatively correct results for the BEC transition.
Second, $\varepsilon_n\neq 0$ components in eq.\ (\ref{Dyson}) are smaller 
than the $\varepsilon_n= 0$ one by $(U\tau)^2$ in the critical region so that they are negligible, 
as seen easily by using the transformation of eq.\ (\ref{COV}).
Thus, the classical-field approximation by Baym {\em et al}.\ \cite{Baym99,Baym99-2} has also been justified by the present consideration.

\section{Results}

\subsection{Numerical procedure}

We explain how to solve eqs.\ (\ref{sigma}) and (\ref{Gamma-eq}) numerically to obtain the reduced self-energy
$\sigma(p)$ and reduced chemical potential $\Gamma$ as a function of the reduced temperature $\tau$. 
Let us introduce the non-interacting correspondent of eq.\ (\ref{chi}) as
\begin{eqnarray}
&&\hspace{-10mm}
\chi^{(0)}(q)\equiv \int\frac{{\rm d}^{3}p}{(2\pi)^3}\frac{1}{(|{\bm p}+{\bm q}/2|^2+1)(|{\bm p}-{\bm q}/2|^2+1)}
\nonumber \\
&&\hspace{1.5mm}
=\frac{\arctan (q/2)}{4\pi q}.
\label{chi^0}
\end{eqnarray}
The second expression has been obtained by (i) performing angular integrations, 
(ii) subsequently expanding $\ln [(p\pm q/2)^2+1]$ in terms of $pq/(p^2+q^2/4+1)$,
(iii) making a change of variables as $p=(q^2/4+1)^{1/2}\tan\theta$ 
and carrying out the $\theta$ integration,
and (iv) comparing the resulting series with the Taylor expansion of $\arcsin x$.
Noting that  $\chi(q)\rightarrow \chi^{(0)}(q)$ for  $q\rightarrow \infty$, we realize that
$\chi(q)\rightarrow 1/8 q$ for $q\gg 1$. 
Hence, it follows that eq.\ (\ref{sigma}) for $p\gg 1$ is well described by
\begin{eqnarray}
&&\hspace{-10mm}
\sigma^{(0)}(p)\equiv \int\frac{{\rm d}^{3}q}{(2\pi)^3} \frac{1}{4\tau^2 q}
\left(\frac{1}{q^2+1}-\frac{1}{|{\bm p}+{\bm q}|^2+1}\right)
\nonumber \\
&&\hspace{1.5mm}
=\frac{\ln (p^2+1)-2+(2/p)\arctan p}{(4\pi\tau)^2 } ,
\label{sigma^0}
\end{eqnarray}
where the second expression has been obtained in the same manner as eq.\ (\ref{chi^0}).

Equations (\ref{sigma}) and (\ref{Gamma-eq}) have been solved iteratively, starting from the non-interacting Green's function and chemical potential in the integrands.
Functions $\chi(q)$ and $\sigma(p)$ are calculated by using 
the integral expressions of $\chi(q)-\chi^{(0)}(q)$ and $\sigma(p)-\sigma^{(0)}(p)$
whose integrands decrease more quickly in the high-momentum region than those of eqs.\ (\ref{chi}) and (\ref{sigma}).
Further, we make a change of variables $p=\sinh u^2$ and $q=\sinh v^2$ for the integrations to cover a wide momentum range 
up to a cutoff momentum $p_{\rm cut}=10^5\sim 10^7$.
We have also stored $\chi(q)$ and $\sigma(p)$ at equal intervals in terms of $v$ and $u$.
These values are used in the next step of iteration with interpolation.
The region ${\bm p}+{\bm q}\sim{\bf 0}$ of eqs.\ (\ref{chi}) and (\ref{sigma}) are handled separately to incorporate more integration points in both the polar and radial integrations.
The convergence has been checked by changing the number of integration points as well as $p_{\rm cut}$.

The above procedure has been repeated for four different approximations with kernels (\ref{v(q)}) and (\ref{v(q)-add})
to check how reliable the predictions on the BEC transition by the present approach are.

\subsection{Results}

\begin{figure}[t]
        \begin{center}
                \includegraphics[width=0.85\linewidth]{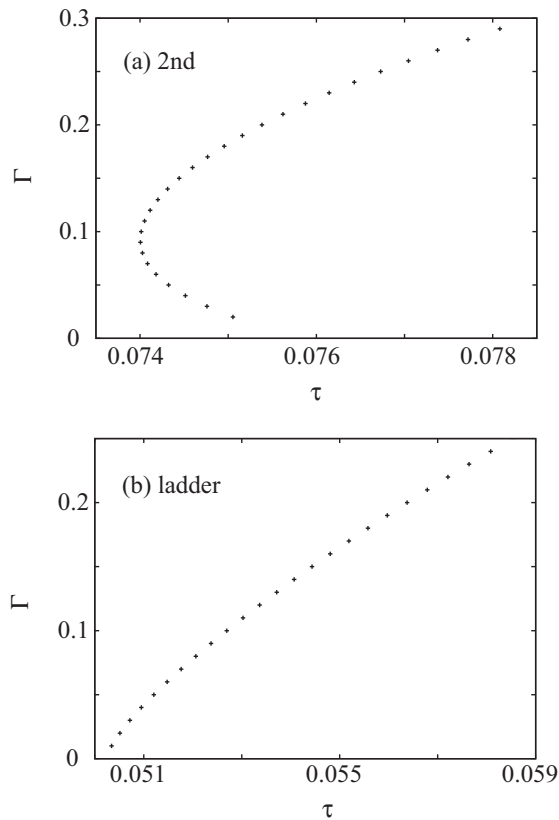}
        \end{center}
        \caption{Reduced chemical potential $\Gamma$ as a function of reduced temperature $\tau$
        in (a) the self-consistent second-order approximation and (b) self-consistent ladder approximation.\label{Fig3}}
\end{figure}

Figure \ref{Fig3} plots reduced chemical potential $\Gamma\propto \Sigma(\vec{0})-\mu$ as a function of reduced temperature $\tau\propto T-1$
in the (a) self-consistent second-order approximation with kernel (\ref{v_2(q)}) and (b) self-consistent ladder approximation
with kernel (\ref{v_pp(q)}).
A continuous BEC transition corresponds to a monotonic decrease of $\Gamma(\tau)$ towards $0$
where BEC is realized.
Thus, Fig.\ \ref{Fig3}(b) from the ladder approximation apparently exhibits a continuous BEC transition 
around $\tau_{\rm c}\!=\! 0.0503$, which translates to $\Delta T_{\rm c}/T_{\rm c}^0\!=\!2.94 a n^{1/3}$
by eqs.\ (\ref{DeltaTc}) and (\ref{c-expression}). 
On the other hand, Fig.\ \ref{Fig3}(a) from the second-order approximation shows a clear sign of a first-order transition
somewhere between $0.074\!\lesssim\! \tau\!\lesssim\! 0.076$ where $\Gamma(\tau)$ is multivalued.
As for the FLEX approximation, we have not even found a solution that approaches $\Gamma\!=\! 0$ continuously; 
here $\Gamma(\tau)$ has a minimum $\Gamma_{\rm min}\approx 0.95$ around $\tau=0.14$.
This strange behavior is brought about by kernel (\ref{v(q)}) that changes sign from negative to positive as $\tau$
is reduced. The same statement holds for the particle-hole approximation of eq.\ (\ref{v_ph(q)}).
Since the present model is quite close to the ideal Bose gas,
we may conclude that the BEC transition is a first-order transition in both the FLEX and particle-hole approximations.

With the diversity of predictions in the present approach, we can hardly say anything definite about the nature of the BEC transition
 with a weak two-body interaction.
However, the fact that some standard $\Phi$'s (i.e., $\Phi_2$, $\Phi_{\rm 2}+\Phi_{\rm ph}$, and $\Phi_{\rm FLEX}$)
exhibit a first-order transition challenges us to clarify unambiguously whether or not
the BEC transition is really continuous.

As for the transition temperature shift, the value $\Delta T_{\rm c}/T_{\rm c}^0\!=\!2.94 a n^{1/3}$
in the present ladder approximation departs from the value $\Delta T_{\rm c}/T_{\rm c}^0\!\simeq \! 2.5 a n^{1/3}$ obtained
by Baym {\em et al}.\, \cite{Baym99-2} with an ultraviolet cutoff $\Lambda$.
However, they also reported that their bubble-sum result $\Delta T_{\rm c}/T_{\rm c}^0\!\simeq \! 1.6 a n^{1/3}$ for the finite $\Lambda$
is increased up to $2.0 a n^{1/3}$ by the extrapolation $\Lambda\rightarrow \infty$. 
Incorporating the same difference $0.4 a n^{1/3}$ into their ladder-summation result yields $\Delta T_{\rm c}/T_{\rm c}^0\!\simeq\! (2.5+0.4) a n^{1/3}$,
which is in good agreement with $2.94 a n^{1/3}$ in the present approach.
If we use $\tau_{\rm c}\!\simeq\! 0.076$ from Fig.\ \ref{Fig3}(a) determined by eq.\ (\ref{Gamma(tau_c)=0}),
we obtain $\Delta T_{\rm c}/T_{\rm c}^0=4.4 a n^{1/3}$ for the second-order approximation, which is also in good agreement with
$\Delta T_{\rm c}/T_{\rm c}^0\!\simeq\! (3.8+0.4) a n^{1/3}$ by Baym {\em et al}. \cite{Baym99-2}
However, it should be noted once again that our second-order result clearly indicates a first-order transition contrary to their assumption.
In their early study, Baym {\em et al}.\ \cite{Baym99} also reported 
$\Delta T_{\rm c}/T_{\rm c}^0\!=\!2.9 a n^{1/3}$ as eq.\ (29),
which apparently agrees with our ladder result mentioned above. 
However, their estimate starts from the second-order
approximation corresponding to our eq.\ (\ref{Sigma_2}) and 
subsequently adopts a trial form $U(k)\approx k^{3/2}k_{\rm c}^{1/2}/[1+(k/k_{\rm c})^{3/2}]$
for $U(k)\equiv 2m[\Sigma_2({\bm k},i\varepsilon_n=0)-\mu]$ at $T=T_{\rm c}$ 
to interpolate between the low- and high-momentum behaviors of 
$\epsilon_{\bm k}+\Sigma({\bm k},i\varepsilon_n=0)-\mu$ with an intermediate parameter $k_{\rm c}$.
Thus, the agreement is accidental and their prefactor $2.9$ should be replaced by 
$3.8$($+0.4$) in their later numerical study for the second-order approximation.\cite{Baym99-2}

\begin{acknowledgments}
This work is supported by a Grant-in-Aid for Scientific Research (C) (No.\ 22540356) from 
the Ministry of Education, Culture, Sports, Science and Technology (MEXT), Japan.
\end{acknowledgments}

\end{document}